\documentclass[prl, twocolumn,superscriptaddress,amsmath,amssymb,showpacs,floatfix,preprintnumbers]{revtex4-2}

\usepackage{amsmath}
\usepackage{graphicx}
\usepackage{dcolumn}
\usepackage{xcolor}
\usepackage{physics}
\usepackage{xr}
\usepackage{siunitx}
\usepackage{tikz}
\usepackage{tabularx}
\usepackage{makecell}
\usepackage{multirow}
\DeclareGraphicsExtensions{.png .jpg .pdf}

\usepackage[normalem]{ulem}

\usepackage{pifont}

\makeatletter
\newcommand*{\addFileDependency}[1]{
  \typeout{(#1)}
  \@addtofilelist{#1}
  \IfFileExists{#1}{}{\typeout{No file #1.}}
}
\makeatother
 
\newcommand*{\myexternaldocument}[1]{%
    \externaldocument{#1}%
    \addFileDependency{#1.tex}%
    \addFileDependency{#1.aux}%
}

\myexternaldocument{si}

\begin{document}

\title{Optical Gain Through Metallic Electro-Optical Effects}

\author{N. Roldan-Levchenko}
\affiliation{School of Physics and Astronomy, University of Minnesota, Minneapolis, Minnesota 55455, USA}
\author{D. J. P. de Sousa}
\affiliation{Department of Electrical and Computer Engineering, University of Minnesota, Minneapolis, Minnesota 55455, USA}
\author{C. O. Ascencio}
\affiliation{School of Physics and Astronomy, University of Minnesota, Minneapolis, Minnesota 55455, USA}
\author{J. D. S. Forte}
\affiliation{Department of Electrical and Computer Engineering, University of Minnesota, Minneapolis, Minnesota 55455, USA}
\author{L. Martin-Moreno}
\affiliation{Instituto de Nanociencia y Materiales de Aragon (INMA), CSIC-Universidad de Zaragoza, 50009 Zaragoza, Spain}
\affiliation{Departamento de F\'{i}sica de la Materia Condensada, Facultad Ciencias, Universidad de Zaragoza, 50009 Zaragoza, Spain}
\author{T. Low}\email{tlow@umn.edu}
\affiliation{Department of Electrical and Computer Engineering, University of Minnesota, Minneapolis, Minnesota 55455, USA}
\affiliation{School of Physics and Astronomy, University of Minnesota, Minneapolis, Minnesota 55455, USA}

\begin{abstract} 
Optical gain is a critical process in today's semiconductor technology and it is most often achieved via stimulated emission. In this theoretical study, we find a resonant TE mode in biased low-symmetry two-dimensional metallic systems which may lead to optical gain in the absence of stimulated emission. We do so by first modeling the optical conductivity using Boltzmann non-equilibrium transport theory and then simulating the scattering problem using a scattered-wave formalism. Assuming that the system may possess a Berry curvature dipole (BCD) and a non-zero Magnetoelectric tensor (MET), we find that the optical conductivity has a non-trivial dependence on the direction of the applied bias, which allows for probing the TE mode. After analyzing the system with one of each of the effects, we find that the resonant TE mode is only accessible when both effects are present. Further studies are necessary to find materials with a suitably large BCD and MET, in order to realize the predictions within this study.

\end{abstract}

\maketitle


\section{I. Introduction}

Optical gain is conventionally understood as the coherent amplification of light resulting from the recombination of electrons in the conduction band with holes in the valence band, a process known as stimulated emission, traditionally requires population inversion~\cite{Samuel2009, Pavesi2000, Low2018}. In this regime, an incoming photon stimulates the emission of additional, phase-coherent photons, leading to an overall increase in the intensity of the transmitted or reflected light. This mechanism forms the basis for laser operation and other active photonic technologies~\cite{Samuel2009, Low2018}. While a number of studies have proposed and experimentally demonstrated gain mechanisms that do not rely on population inversions, the majority of these approaches still fundamentally involve stimulated emission as the key amplification process~\cite{Harris1989, Imamolu1994, Mysyrowicz2019, Doronin2019}. More recently, metallic electro-optic (EO) effects have been predicted to enable intriguing optical responses in non-centrosymmetric systems, including the possibility of non-reciprocal optical gain~\cite{PhysRevLett.130.076901, PhysRevB.109.245126, PhysRevB.110.115421, PhysRevB.107.125151, PhysRevApplied.22.L041003}. The gain mechanism is fundamentally distinct from stimulated emission-based processes, as they originate purely from non-trivial intraband dynamics of Bloch electrons in the presence of static fields, without relying on interband transitions, representing a new paradigm of light amplification with potential applications in active photonic and optoelectronic devices, particularly in the terahertz regime.


While metallic EO effects were originally ascribed to the presence of a finite Berry curvature dipole (BCD) of Bloch electrons on the Fermi surface~\cite{PhysRevLett.130.076901, PhysRevB.107.125151}, recent developments have uncovered alternative intraband mechanisms that could enable similar non-reciprocal responses in quantum materials~\cite{PhysRevB.110.115421, PhysRevB.110.125401, Ma2025}. A notable example is the linear magnetoelectric EO effect, which arises from the magnetic moment texture of Bloch electrons on the Fermi surface~\cite{PhysRevB.110.115421}. Owing to the intimate relation between the magnetic moment and Berry-curvature textures, a comprehensive description of the non-reciprocal optical gain must incorporate effects stemming from both features.

In this work, we conduct a comprehensive investigation of the influence of metallic EO effects on non-reciprocal optical gain in non-centrosymmetric quantum materials. We adopt a unified theoretical framework that captures the emergence of EO responses originating from the BCD and the orbital magnetic moment of Bloch electrons at the Fermi surface~\cite{PhysRevB.110.115421}. Within this formalism, we derive the full set of quantum transport coefficients characterizing bias-induced metallic EO effects and implement a numerical electromagnetic wave scattering method to compute the resulting optical response. This approach enables a systematic analysis of optical gain optimization, including mapping the gain phase space in terms of key material parameters such as intrinsic anisotropy. Our results reveal that the recently proposed magnetoelectric EO effect~\cite{PhysRevB.110.115421} is governed by a non-Hermitian magnetoelectric tensor (MET) and gives rise to gain features that are qualitatively distinct from those induced by the BCD. We compare and contrast these two mechanisms, highlighting their individual and combined roles in shaping the non-reciprocal response. A key finding of this work is the central role played by the transverse-electric (TE) mode in enhancing gain driven by metallic EO effects. In particular, we show that the simultaneous presence of MET- and BCD-induced EO responses enables closer proximity to the TE-resonance with respect to frequency, thereby unlocking a pathway to maximize gain. Our results establish a general strategy for engineering non-reciprocal optical amplification via metallic EO effects, with implications for terahertz photonics. 

The paper is organized as follows. In Section II, we introduce the theoretical framework employed in this work, including the classification of the mechanisms responsible for metallic EO effects and the electromagnetic wave scattering methodology used to compute the optical response. Section III presents the main results, starting with a discussion of the role of the TE mode resonance in subsection A. Subsection B focuses on the optical gain induced solely by the BCD, while subsection C examines the contribution from the MET EO effect. In subsection D, we analyze the interplay between these two mechanisms and their combined impact on optical gain. Finally, Section IV provides a summary and concluding remarks.

\section{II. Formalism}
\label{sec:Form}
We start by reviewing the formalism used in this study, beginning with the generalized constitutive relations under applied bias, followed by an overview of the scattering mechanism framework.

\subsection{A. Constitutive relations in the presence of bias}

We focus on the simplest case where the two-dimensional (2D) electron system is described by a gapped Dirac Hamiltonian (see Appendix A for details). In the low-frequency limit, $\hbar \omega \ll \Delta$, with $\Delta$ being the optical gap, the system's electromagnetic response is dominated by intraband transitions. In this regime, the constitutive relation takes the form $J_0^{\beta}(\omega) = \sigma^{\alpha\beta}_{\textrm{Drude}}(\omega) E^{\beta}_{\omega}$, which captures the conventional AC Drude response. The corresponding transport coefficient is explicitly given by:
\begin{eqnarray}
\sigma_{\textrm{Drude}}^{\alpha\beta}(\omega) = \displaystyle \frac{e^2\tau}{1 - i\omega\tau}\sum_n \int \frac{d^2\textbf{k}}{(2\pi)^2}\left(-\frac{\partial f_{n\textbf{k}}^0}{\partial \epsilon_{n\textbf{k}}}\right)v_{n\textbf{k}}^{\alpha}v_{n\textbf{k}}^{\beta}, 
    \label{Eq2}
\end{eqnarray}
where $f^0_{n\textbf{k}}$ is the fermi-dirac distribution, $\epsilon_{n\textbf{k}}$ is the electron dispersion with associated Bloch velocity $v^{\alpha}_{n\textbf{k}} = (1/\hbar)\partial \epsilon_{n\textbf{k}}/\partial k_{\alpha}$, and $\tau = 1/\gamma$ is the relaxation time. Since the dilute impurity limit requires $\gamma \ll \omega$, the semiclassical approach adopted here confines the frequency range to $\gamma \ll \omega \ll \Delta/\hbar$.

For our toy model system, the AC drude conductivity tensor, in the basis $\textbf{E}_{\omega} = E^x_{\omega}\hat{\textbf{x}} + E^y_{\omega}\hat{\textbf{y}}$, is fully diagonal $\boldsymbol{\sigma}_{\textrm{Drude}}(\omega) = \textrm{diag}[\sigma^{xx}_{\textrm{Drude}}(\omega) \ \ \sigma^{yy}_{\textrm{Drude}}(\omega)]$ and isotropic, i.e., $\sigma^{xx}_{\textrm{Drude}}(\omega) = \sigma^{yy}_{\textrm{Drude}}(\omega) = \sigma_{\textrm{Drude}}(\omega)$, with
\begin{eqnarray}
       \sigma_{\textrm{Drude}}(\omega) = \frac{2e^2}{h}\frac{\omega_F}{\gamma - i\omega}\left[1 - \left(\frac{\omega_{\Delta}}{\omega_F}\right)^2\right].
       \label{eq:drude}
\end{eqnarray}
Here, $\omega_F = \mu/\hbar$ corresponds to the Fermi level $\mu$, while $\omega_{\Delta} = \Delta/2\hbar$ defines a characteristic frequency associated with the energy gap. In a later section, we will consider the case of slight anisotropy, where the components $\sigma^{xx}_{\textrm{Drude}}$ and $\sigma^{yy}_{\textrm{Drude}}$ may differ. In these cases, we will quantify the anisotropy by $\eta = \sigma^{yy}_{\textrm{Drude}}/\tilde{\sigma}^{yy}_{\textrm{Drude}}$. 

The presence of a static electric field, $\textbf{E}_0$, alters the system’s electromagnetic response by introducing a correction to the constitutive relation, leading to electro-optical (EO) effects~\cite{PhysRevB.110.115421}:
\begin{eqnarray}
J_0^{\beta}(\omega) \rightarrow J_0^{\beta}(\omega) + J_{\textrm{EO}}^{\beta}(\omega),
    \label{EOeffects}
\end{eqnarray}
where $J_{\textrm{EO}}^{\beta}(\omega)$ can be written most generally as a function of the electromagnetic fields, ($\textbf{E}_{\omega}$, $\textbf{B}_{\omega}$): 
\begin{eqnarray}
J_{\textrm{EO}}^{\alpha}(\omega) = \sigma^{\alpha\beta}_{\textrm{D}}(\omega) E^{\beta}_{\omega} + \sigma^{\alpha\beta}_{\textrm{G}}(\omega) B^{\beta}_{\omega},
    \label{EOeffects2}
\end{eqnarray}
in time-reversal symmetric systems lacking inversion symmetry~\cite{PhysRevB.110.115421}. In contrast to previous studies that considered only the contribution of $\sigma^{\alpha\beta}_{\textrm{D}}(\omega)$~\cite{PhysRevLett.130.076901, PhysRevApplied.22.L041003, PhysRevB.109.245126}, this work also incorporates bias-induced magnetoelectric terms, $\sigma^{\alpha\beta}_{\textrm{G}}(\omega)$, and investigates their combined role in enabling optical gain. This leads to a more comprehensive description of the system's EO response. In what follows, we analyze the nature of the bias-induced corrections in more detail.


\subsubsection{A1. EO effects derived from the Berry curvature dipole}

We begin by examining the nature of the metallic EO effect described by the $\sigma^{\alpha\beta}_{\textrm{D}}(\omega)$ coefficient. Prior studies have shown that this contribution originates from the BCD of Bloch electrons~\cite{PhysRevLett.130.076901, PhysRevApplied.22.L041003, PhysRevB.109.245126, PhysRevB.110.115421}. To linear order in the optical fields, the applied bias introduces two primary corrections to the intraband optical conductivity, captured by the following terms
\begin{eqnarray}
\sigma_{D}^{\alpha\beta}(\omega) =  (\epsilon_{\alpha\gamma\lambda}\chi^{\lambda\beta}(\omega) - \epsilon_{\alpha\beta\lambda}\tilde{\chi}^{\lambda\gamma})E_0^{\gamma}, 
    \label{Eq10}
\end{eqnarray}
where the two coefficients are related according to
\begin{subequations}
\begin{eqnarray}
\chi^{\alpha\beta}(\omega) = \displaystyle \frac{\tilde{\chi}^{\alpha\beta}}{1 - i\omega\tau},
    \label{secIII.eq7a}
\end{eqnarray}
\begin{eqnarray}
\tilde{\chi}^{\alpha\beta} = \displaystyle \frac{e^3\tau}{\hbar} \sum_n \int \frac{d^2\textbf{k}}{(2\pi)^2}\left(-\frac{\partial f_{n\textbf{k}}^0}{\partial \epsilon_{n\textbf{k}}}\right)\Omega_{n\textbf{k}}^{\alpha}v_{n\textbf{k}}^{\beta}. 
    \label{secIII.eq7b}
\end{eqnarray}
\end{subequations}

In the above, $\boldsymbol{\Omega}_{n\mathbf{k}}=-\text{Im}\bra{\nabla_{\mathbf{k}}u_{n\mathbf{k}}} \times \ket{\nabla_{\mathbf{k}}u_{n\mathbf{k}}}$ is the Berry curvature associated with the Bloch state $\ket{u_{n\mathbf{k}}}$~\cite{PhysRevB.110.115421}. We also note that the momentum space integral in Eq.~(\ref{secIII.eq7b}) is simply the BCD up to a $\hbar$ factor. For instance, integration by parts shows that
\begin{eqnarray}
\displaystyle \hbar \int \frac{d^2\textbf{k}}{(2\pi)^2}\left(-\frac{\partial f_{n\textbf{k}}^0}{\partial \epsilon_{n\textbf{k}}}\right)\Omega_{n\textbf{k}}^{\alpha}v_{n\textbf{k}}^{\beta} = \int \frac{d^2\textbf{k}}{(2\pi)^2}f_{n\textbf{k}}^0 \frac{\partial \Omega_{n\textbf{k}}^{\alpha}}{\partial k_{\beta}}. \nonumber \\
    \label{Eq12}
\end{eqnarray}

\begin{figure}[t]
\includegraphics[width = \linewidth]{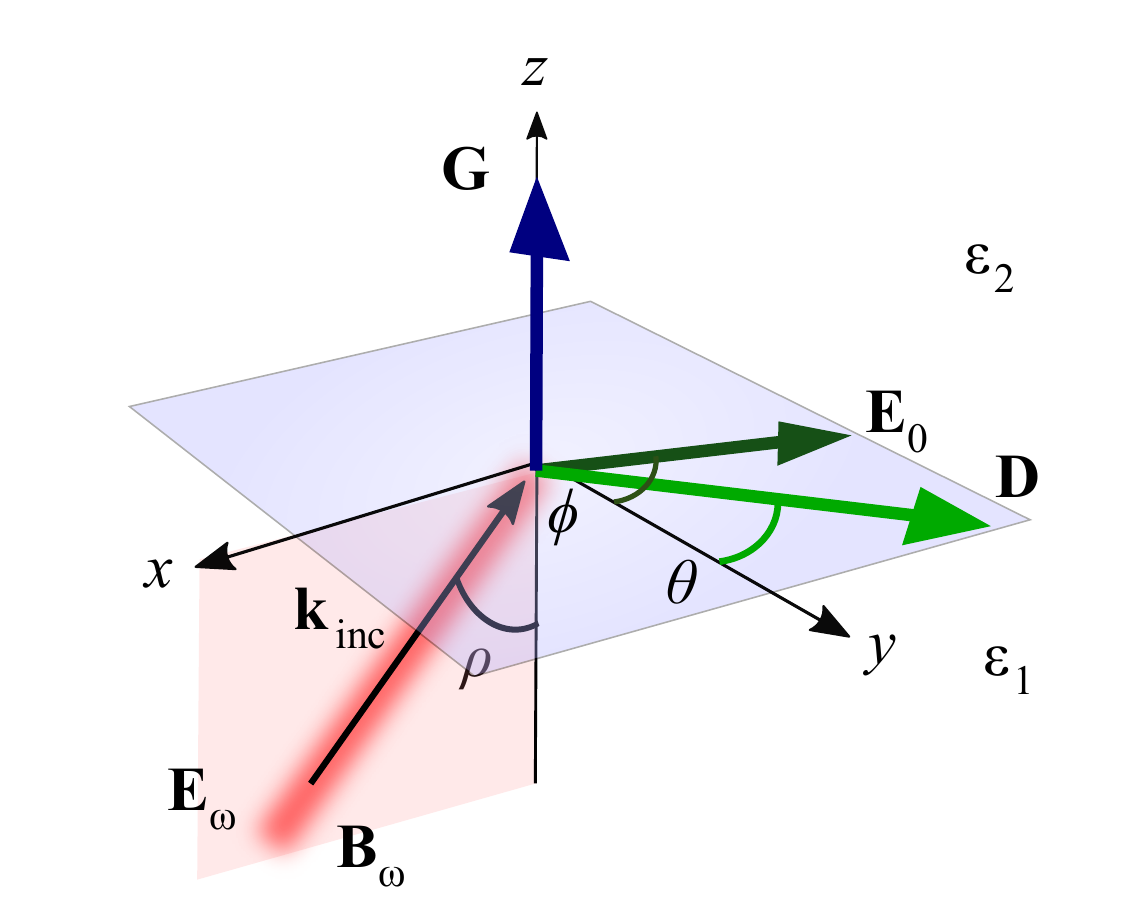}
\caption{Schematic and definition of parameters of the scattering formalism alongside the directions of the non-zero components of the $\textbf{G}$ tensor and $\textbf{D}$ dipole of a 2D metal at $z = 0$. While the $\textbf{G}$ tensor is fixed in the out-of-plane direction, the direction of $\textbf{D}$ ($\textbf{E}_0$) in the $xy$ plane is determined by the angle $\theta$ ($\phi$), measured with respect to the $y$ axis. The 2D metal is in between dielectrics with relative permittivity $\epsilon_1$ ($z < 0$) and $\epsilon_2$ ($z > 0$). We assume that the plane of incidence is the $xz$ plane, containing the incidence wavevector $\textbf{k}_{\textrm{inc}}$.} 
\label{fig:schem}
\end{figure}

Such metallic EO correction to the optical conductivity can be fully expressed in a compact form by introducing the BCD tensor (up to the $\hbar$ factor) $\textbf{D} = \sum_{n\textbf{k}}(\partial f^0_{n\textbf{k}}/\partial \epsilon_{n\textbf{k}})\textbf{D}_{n\textbf{k}}$ (In this work, we will use $\textbf{D}$ to refer exclusively to the Berry dipole tensor. Thus, it should not be confused with the displacement electric field), with components $D^{\alpha\beta}_{n\textbf{k}} = v^{\alpha}_{n\textbf{k}}\Omega^{\beta}_{n\textbf{k}}$, as
\begin{eqnarray}
\boldsymbol{\sigma}_{\textrm{D}}(\omega)  = \displaystyle \frac{e^3}{\hbar}\left(-\frac{1}{\gamma}\textbf{D} \cdot \textbf{E}_0+\frac{1}{\gamma - i\omega}\textbf{F}_0 \cdot \textbf{D}\right),
    \label{Eq13}
\end{eqnarray}
where we have defined a fully antisymmetric electric field tensor $\textbf{F}_0$, with components $F_0^{\alpha\beta} = -\epsilon_{\alpha\beta\gamma}E_0^{\gamma}$. In previous studies, these contributions were referred to as Hermitian (H) and non-Hermitian (NH) EO effects, respectively~\cite{PhysRevLett.130.076901, PhysRevApplied.22.L041003, PhysRevB.109.245126, PhysRevB.110.115421}. In this work, we assume the setup illustrated in Fig.~\ref{fig:schem}, where the system supports a single BCD component oriented along an in-plane direction determined by an angle $\theta$ in relation to the $y$-axis. Assuming that the orientation of the static electric field $\textbf{E}_0$ make a angle $\phi$ with the same axis, the explicit forms for the Hermitian and non-Hermitian BCD-induced conductivity tensors are
\begin{eqnarray}
    \boldsymbol{\sigma}_{\textrm{D}}^{\textrm{H}} = -\frac{e^2}{\pi\hbar}\frac{\xi}{\gamma} 
    \begin{bmatrix}
        0 & -\cos(\theta - \phi) \\
        \cos(\theta - \phi) & 0 
    \end{bmatrix},
    \label{eq:Hermitian}
\end{eqnarray}
and
\begin{eqnarray}
    \boldsymbol{\sigma}_{\textrm{D}}^{\textrm{NH}} = \frac{e^2}{\pi\hbar}\frac{\xi}{\gamma - i\omega}
    \begin{bmatrix}
        -\sin\theta\cos\phi & \cos\theta\cos\phi \\
        -\sin\theta\sin\phi & \cos\theta\sin\phi
    \end{bmatrix}. 
    \label{eq:bnh}
\end{eqnarray}
Here, $\xi = \pi e E_0 D_0 / \hbar$, where $E_0 = |\textbf{E}_0|$ and $D_0$ denotes the magnitude of the BCD [see Appendix B for further details]. Notably, the angular degrees of freedom $\theta$ and $\phi$ offer control over the structure of the conductivity tensor, a feature that has not been emphasized in previous works~\cite{PhysRevLett.130.076901, PhysRevApplied.22.L041003}. In particular, the Hermitian component of the response can be completely suppressed by orienting the bias perpendicular to the BCD, i.e., $\boldsymbol{\sigma}_{\textrm{D}}^{\textrm{H}} = 0$ when $\theta - \phi = \pi/2$, for any $D_0$ and $E_0$. In subsequent sections, we will explore this freedom to enable favorable conditions for maximizing optical gain.

Next, we summarize the EO effect contribution that couples to the $\textbf{B}_{\omega}$ optical field.

\subsubsection{A2. EO effects derived from the magnetic moment texture}
\label{subsubsec:A1}

In this section, we examine the nature of the metallic EO effect described by the $\sigma^{\alpha\beta}_{\textrm{G}}(\omega)$ coefficient. A prior study revealed that the interplay between the Berry curvature and magnetic moment texture of Bloch electrons provides a mechanism to enable a $\textbf{E}_0$-induced response given by~\cite{PhysRevB.110.115421}
\begin{eqnarray}
\sigma_{G}^{\alpha\beta}(\omega) = \epsilon_{\alpha\gamma\lambda}\zeta^{\lambda\beta}(\omega)E_0^{\gamma},
    \label{secII.eq3}
\end{eqnarray}
where, explicitly
\begin{eqnarray}
\zeta^{\alpha\beta}(\omega) = \displaystyle \frac{e^2}{\hbar} \frac{i\omega\tau}{i\omega\tau - 1}\sum_n \int \frac{d^2\textbf{k}}{(2\pi)^2}\left(-\frac{\partial f_{n\textbf{k}}^0}{\partial \epsilon_{n\textbf{k}}}\right)\Omega_{n\textbf{k}}^{\alpha}m_{n\textbf{k}}^{\beta}, \nonumber \\
    \label{secII.eq4}
\end{eqnarray}

and $\mathbf{m}_{n\mathbf{k}}$ is the total magnetic moment of Bloch electrons~\cite{PhysRevB.110.115421}.

Similar to the EO induced by the BCD, it is possible to rewrite the above contribution in a compact form by introducing the EO MET $\textbf{G} = \sum_{n\textbf{k}}(-\partial f^0_{n\textbf{k}}/\partial \epsilon_{n\textbf{k}})\textbf{G}_{n\textbf{k}}$, with components related to $G^{\alpha\beta}_{n\textbf{k}} = \Omega^{\alpha}_{n\textbf{k}}m^{\beta}_{n\textbf{k}}$. This results in
\begin{eqnarray}
\boldsymbol{\sigma}_{\textrm{G}}(\omega)  = \displaystyle \frac{e^2}{\hbar}\frac{i\omega}{i\omega - \gamma}\textbf{F}_0 \cdot \textbf{G}.
    \label{Eq9}
\end{eqnarray}
We note that such contribution is non-Hermitian and, therefore, it might also enable optical gain. Note that the current response now explicitly depends on the transmitted (incidence) angle $\rho_2$ ($\rho$). Due to our assumption that we have a 2D system, $\textbf{G}$ has a single component oriented along the $z$-axis, $G^{zz}$ (associated with $\Omega^{z}_{n\textbf{k}}m^{z}_{n\textbf{k}}$). Thus, only $B_{\omega}^z$ couples to $m^{z}_{n\textbf{k}}$. In subsequent section, $\boldsymbol{\sigma}_{\textrm{G}}(\omega)$ is re-written in the $\textbf{E}_{\omega}$-field basis, from which we obtain a dependence on $\rho_2$. For a detailed discussion, see Appendix C. 

The explicit shape of the magnetoelectric EO contribution is
\begin{eqnarray}    
    \boldsymbol{\sigma}_{\textrm{G}}(\omega) = \frac{e^2}{\hbar}\frac{i\omega\Gamma}{\gamma - i\omega}
    \begin{bmatrix}
        0 & -\sin\rho_2\cos\phi \\
        0 & -\sin\rho_2\sin\phi
    \end{bmatrix},
    \label{eqsigma_G}
\end{eqnarray}
where, similar to the previous section, $\Gamma = \epsilon_2 E_0G^{zz} /c$, $\epsilon_2$ is the permittivity of the surrounding media and $c$ is the vacuum speed of light. In writing Eq.~\ref{eqsigma_G} we have also assumed that the choice of $xyz$-coordinate orientations are dependent on where the optical field is incident. Here, we chose the coordinates such that the optical field is always incident in the $xz$-plane, i.e. $k_y=0$ (the $y$-component of the wavevector $\mathbf{k}$). In the previous section, however, the coordinate orientations were determined by the crystal axis, i.e. which way the BCD pointed. We will further discuss this important distinction when analyzing our resulting findings.

 Together, Eqs.~(\ref{eq:drude}), (\ref{eq:Hermitian}), (\ref{eq:bnh}) and (\ref{eqsigma_G}), account for the relevant optical responses arising from the coupling between optical fields, static bias and the wave function of Bloch electrons in inversion-broken time-reversal symmetric systems. Next, we describe the electromagnetic wave scattering approach employed in this work.


\subsection{B. The Electromagnetic wave scattering problem}

In the following, we describe the formalism adopted in this work for solving the scattering problem of s- and p-polarized electromagnetic waves through a 2D material, with a given associated optical conductivity tensor $\boldsymbol{\sigma}(\omega)$, sandwiched in between dielectric media with refractive indices $n_1$ and $n_2$. This formalism closely follows that presented in Ref.~\cite{oliva}. 

\subsubsection{B1. Oblique incidence}

Without loss of generality, we assume that the wave is incident in the $xz$-plane and traveling in the $+z$ direction. For clarity, the phase factors $e^{-i\omega t}$ are omitted. Thus, the incident, reflected, and transmitted components of the optical fields, $\textbf{E} \equiv \textbf{E}_{\omega}$ and $\textbf{B} \equiv \textbf{B}_{\omega}$ in this section, can be written as~\cite{oliva}:
\begin{equation}
    \textbf{E}_{\textrm{inc}} = 
        \begin{bmatrix}
                -a_p\cos{\rho_1} \\
                a_s \\
                a_p\sin{\rho_1} \\
        \end{bmatrix}
    \ \ \textbf{B}_{\textrm{inc}} = 
        \frac{n_{1}}{z_{0}}\begin{bmatrix}
                -a_s\cos{\rho_1} \\
                -a_p \\
                a_s\sin{\rho_1}
        \end{bmatrix}, 
        \label{inc}
\end{equation}

\begin{equation}
    \textbf{E}_{\textrm{ref}} = 
        \begin{bmatrix}
                r_p\cos{\rho_1} \\
                r_s \\
                r_p\sin{\rho_1} \\
        \end{bmatrix} 
    \ \ \textbf{B}_{\textrm{ref}} = 
        \frac{n_{1}}{z_{0}}\begin{bmatrix}
                r_s\cos{\rho_1} \\
                -r_p \\
                r_s\sin{\rho_1}
        \end{bmatrix},
        \label{ref}
\end{equation}

\begin{equation}
    \textbf{E}_{\textrm{tr}} = 
        \begin{bmatrix}
                -t_p\cos{\rho_2} \\
                t_s \\
                t_p\sin{\rho_2} \\
        \end{bmatrix}
    \ \ \textbf{B}_{\textrm{tr}} = 
        \frac{n_{2}}{z_{0}}\begin{bmatrix}
                -t_s\cos{\rho_2} \\
                -t_p \\
                t_s\sin{\rho_2}
        \end{bmatrix},
        \label{tr}
\end{equation} 
respectively, where $a$, $r$, $t$ are the incident, reflected, and transmitted waves amplitudes, respectively; the subscripts $s$ and $p$ denote the s-polarized (TE) and p-polarized (TM) components of the wave; $\rho_1$ and $\rho_2$ are the angles of incidence and transmission; and $z_0$ is the vacuum impedance. Lastly, $\rho_1$ is related to $\rho_2$ by Snell's law, i.e. $n_1\sin{\rho_1} = n_2\sin{\rho_2}$. The setup is schematically shown in Fig.~\ref{fig:schem}. \\

The boundary conditions which relate the fields on the top and bottom dielectrics are:
\begin{equation}
    E^{x,y}_{\textrm{tr}} = E^{x,y}_{\textrm{inc}} + E^{x,y}_{\textrm{ref}},
    \label{b1}
\end{equation}
\begin{equation}
    B^{x}_{\textrm{tr}} - B^{x}_{\textrm{inc}} - B^{x}_{\textrm{ref}} = \sigma^{yx}(\omega)E^{x}_{\textrm{tr}} + \sigma^{yy}(\omega)E^{y}_{\textrm{tr}},
    \label{b2}
\end{equation}
and
\begin{equation}
    B^{y}_{\textrm{tr}} - B^{y}_{\textrm{inc}} - B^{y}_{\textrm{ref}} = -\sigma^{xx}(\omega)E^{x}_{\textrm{tr}} - \sigma^{xy}(\omega)E^{y}_{\textrm{tr}}.
    \label{b3}
\end{equation}

Combining in Eqs.~(\ref{inc})-(\ref{tr}) with Eqs.~(\ref{b1})-(\ref{b3}), and solving for $t_s$,$t_p$,$r_s$,$r_p$ in terms of $a_s$, $a_p$, we obtain:
\begin{equation}
    t_s = \frac{2n_1\alpha\cos{\rho_{1}}}{\lambda}a_s + \frac{2n_1\kappa\cos{\rho_{1}}}{\lambda}a_p,
    \label{ts}
\end{equation}
\begin{equation}
    t_p = \frac{2n_1\eta\cos{\rho_{1}}}{\lambda}a_s + \frac{2n_1\beta\cos{\rho_{1}}}{\lambda}a_p,
    \label{tp}
\end{equation}
\begin{equation}
    r_s = \left(\frac{2n_1\alpha\cos{\rho_{1}}}{\lambda} - 1\right)a_s + \frac{2n_1\kappa\cos{\rho_{1}}}{\lambda}a_p,
    \label{rs}
\end{equation}
and
\begin{equation}
    r_p = -\frac{2n_1\eta\cos{\rho_{2}}}{\lambda}a_s + \left(1 - \frac{2n_1\beta\cos{\rho_{2}}}{\lambda}\right)a_p,
    \label{rp}
\end{equation}
where $\alpha = n_1\cos{\rho_2} + n_2\cos{\rho_1} + z_0\sigma^{xx}\cos{\rho_2}\cos{\rho_1}$, \\ $\eta = z_0\sigma^{xy}\cos{\rho_2}$, $\kappa = z_0\sigma^{yx}\cos{\rho_{1}}$, $\beta = n_1\cos{\rho_1} + n_2\cos{\rho_2} + z_0\sigma^{yy}$, and $\lambda = \alpha\beta - \eta\kappa$. Given the final transmittance coefficients written above, we briefly discuss the condition for optimizing the transmission in the following. 

\subsubsection{B2. Optimization condition for the transmission}

The Hermitian transmittance matrix $\mathbf{T}$, which relates the intensity of the incident wave to that of the transmitted wave, given in the s- and p-polarized basis, is $\mathbf{T} = (n_2 \cos{\rho_{2}}/n_1 \cos{\rho_{1}})\mathbf{t}^{\dagger}\mathbf{t}$, where~\cite{hecht}
\begin{equation}
    \mathbf{t} = 
    \begin{bmatrix}
        t_{ss} & t_{sp} \\
        t_{ps} & t_{pp}
    \end{bmatrix}.
\end{equation}
Diagonalization of the matrix $\mathbf{T}$ yields two real eigenvalues, denoted $T_{\min}$ and $T_{\max}$, with $T_{\max} \geq T_{\min}$. The incident wave, composed of s- and p-polarized components, that maximizes the transmittance corresponds to the eigenvector associated with the largest eigenvalue, $T_{\max}$. A detailed derivation of this result is provided in the Appendix D.


\section{III. Results and Discussions}

We now present our main results. We begin by analyzing the conditions under which the TE mode can exist and propagate in the presence of biased metallic systems. Subsequently, we discuss our main numerical results, organized as follows: We begin by studying the optical gain mediated by the BCD (section VI. B) and magnetic moment texture (section VI. C) of Bloch electrons at the Fermi surface individually. Then, we analyze their simultaneous impact (section VI. D). 

To this end, we consider a generic 2D metal at $z = 0$, placed between two dielectric media with $\epsilon_1 = \epsilon_2 = 1$. Next, we apply a static, in-plane bias $\textbf{E}_0$ to the metal and consider an optical field $\textbf{E}_{\omega}$ traveling in the $+z$-direction in the $xz$-plane at an incidence angle $\rho = \rho_1 = \rho_2$, as shown in Fig.~\ref{fig:schem}. For the base electronic structure model, we assume a band-gap energy $\Delta = 5~\text{meV} \approx 1.21~\text{THz}$ and a Fermi energy within the conduction band, $\mu = 5.1~\text{meV}$. Here, we consider $\gamma = 10^{11}~\text{rad/s} \approx 0.02~\text{THz}$. Thus, our analysis holds for frequencies within the interval $0.02 \ll \omega / (2\pi) \ll 1.21~\text{THz}$. Although the following analysis seeks to elucidate the coupling between $\textbf{E}_{\omega}$ and $\textbf{E}_0$ via the metallic EO effects discussed earlier, we focus our attention on the regime where the transmittance of $\textbf{E}_{\omega}$ is above unity. For the remainder of the paper, we use the terms ‘transmissive gain’, ‘optical gain’, and ‘gain’ interchangeably to mean the same phenomenon.

\subsection{A. TE-Mode Resonance}
We begin by analyzing the role of the TE mode resonance in the transmittance response and its potential connection with the optical-gain.

Our findings indicate that a resonant TE mode can be approached in a 2D metallic system possessing both a finite BCD and a sizable magnetic moment on the Fermi surface, through appropriate tuning of the external bias field $\textbf{E}_0$. To show this, we analyze the poles of $t_{ss}$ explicitly. Assuming that the optical conductivity tensor can be tuned to a diagonal and anisotropic form, a condition that we demonstrate to be achievable in subsequent sections, the transmission coefficient for a TE-polarized incident wave, under the assumption $n_1 = n_2 = 1$, given by Eq.~(\ref{ts}), reads
\begin{equation}
    t_{ss} = \frac{2\cos{\rho}}{2\cos{\rho} + z_0\sigma^{yy}}.
    \label{grand}
\end{equation}
Therefore, the TE-polarized optical field can excite a resonant mode when the conditions $2\cos\rho + z_0\text{Re}(\sigma^{yy}) = 0$ and $\text{Im}(\sigma^{yy}) = 0$ are simultaneously satisfied.

We emphasize that this mode is within the light cone, since we assume scattering states. Hence, $\cos{\rho}$ is real and $\rho \in [0, \pi/2)$. In the ideal limit, $2\cos\rho + z_0\text{Re}(\sigma^{yy}) = 0 \rightarrow \text{Re}(\sigma^{yy}) = -(2/z_0) \cos \rho$. But because $\rho \in [0, \pi/2) \rightarrow \cos\rho > 0$, we find that the resonant condition requires $\text{Re}(\sigma^{yy}) < 0$, which implies optical gain. Qualitatively, the losses are compensated by the resonant energy exchange between the in-plane bias and the optical field.  


Our study finds that optical gain is maximized as one approaches the resonance. In particular, this resonance corresponds to a TE mode, which is typically dismissed in the analysis of light–matter interaction in 2D systems, where bound TM modes are favored due to their enhanced field confinement~\cite{goncalves, Low2016, Menabde2016}. Here, however, the TE resonance plays a central role, becoming crucial to enable high optical gain, a feature not recognized in previous works. In a subsequent section, we will demonstrate that while optical gain can arise in systems exhibiting either BCD-induced or magnetoelectric EO effects, access to the resonant mode requires the simultaneous presence of both effects.

\begin{figure}[t]
\includegraphics[width = 0.95\linewidth]{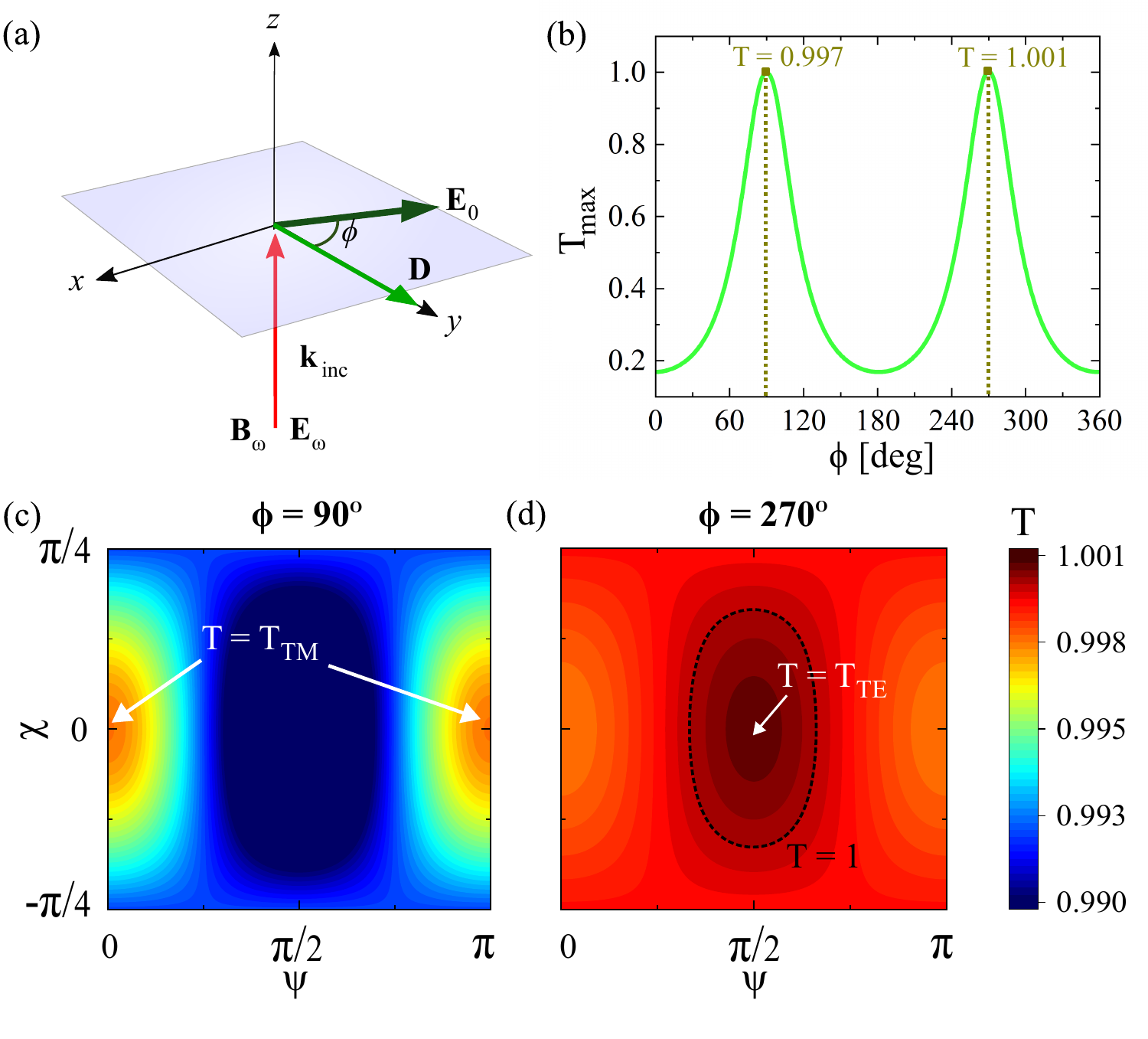}
\caption{Normal scattering in the presence of only $\textbf{D}$, showing that TE-polarized waves and $\mathbf{D}\perp \mathbf{E}_{\textrm{0}}$ with a specific orientation maximizes optical gain. (a) Schematic including only $\mathbf{D}$ and $\mathbf{E}_{\textrm{0}}$. The optical field is assumed to be normally incident ($\rho = 0^{\circ}$); thus, we also fix the direction of $\mathbf{D}$ to point along +$y$ ($\theta = 0^{\circ}$). (b) Transmittance as a function of $\textbf{E}_{\textrm{0}}$ direction with respect to the $y$-axis at the optimal polarization ($T = T_{\textrm{max}}$). (c)-(d) Transmittance of a normally incident-wave with the polarization determined by the tilt ($\Psi$) and ellipticity ($\chi$) angles. In panels (b)-(d), $\rho = 0^{\circ}$, $\theta = 0^{\circ}$, $\omega/(2\pi) = 0.5$ THz, $\epsilon_1 = \epsilon_2 = 1$, ${D}_{\textrm{0}} = 40$ nm, and ${E}_{\textrm{0}} = 8 \times 10^4$ V$\cdot$m$^{-1}$.}
\label{fig:just}
\end{figure}


\subsection{B. Optical Gain through BCD-induced electro-optic effects}

Here, we focus on the optical gain induced solely by the BCD. As demonstrated in Refs.~\cite{PhysRevLett.130.076901, PhysRevApplied.22.L041003}, while the Hermitian component of the EO effect, $\boldsymbol{\sigma}_{\textrm{D}}^{\textrm{H}}$, is non-dissipative, the non-Hermitian part, $\boldsymbol{\sigma}_{\textrm{D}}^{\textrm{NH}}$, can lead to negative power dissipation for a given optical field chirality, i.e., it is this NH contribution that drives optical gain. 
To further investigate this effect, we exploit the freedom introduced by $\phi$ and the polarization of the optical field, to identify the conditions that maximize transmission of normally incident light ($\rho = 0^\circ$) and, consequently, optical gain. Since we are considering only normally-incident light, we fix the direction of $\textbf{D}$ along +$\hat{\textbf{y}}$ ($\theta = 0^{\circ}$).

Figure~\ref{fig:just}(b) displays the transmission as a function of the in-plane orientation of the static bias field $\textbf{E}_0$ (i.e., angle $\phi$). The results reveal that maximum transmission occurs when $\textbf{D} \perp \textbf{E}_0$, corresponding to a suppressed Hermitian EO response ($\boldsymbol{\sigma}_{\textrm{D}}^{\textrm{H}} = 0$). Here, the transmission maxima at $\phi = 90^\circ$, $270^\circ$ correspond to the situation when the anisotropy of the effective conductivity, including the Drude contribution and the BCD-induced EO effect, is maximized. That is, given the setup  $\textbf{D} \perp \textbf{E}_0$, the total effective conductivity assumes the form $\boldsymbol{\sigma}_{\textrm{eff}}(\omega) = \textrm{diag}[\sigma_{\textrm{eff}}^{xx}(\omega) \ \ \sigma_{\textrm{eff}}^{yy}(\omega)]$, with $|\sigma_{\textrm{eff}}^{xx}(\omega) - \sigma_{\textrm{eff}}^{yy}(\omega)|$ being maximized. The explicit forms of the component of the effective conductivity are $\sigma_{\textrm{eff}}^{xx}(\omega) = \sigma_{\textrm{Drude}}(\omega) $ and $\sigma_{\textrm{eff}}^{yy}(\omega) = \sigma_{\textrm{Drude}}(\omega)  - \sigma_{\textrm{D}}^{\textrm{NH};yy}(\omega)$, where $\sigma_{\textrm{D}}^{\textrm{NH};yy}(\omega) = (e^2/\pi\hbar) \xi/(\gamma - i\omega)$. Henceforth, $\boldsymbol{\sigma}_{\textrm{eff}}$ and $\boldsymbol{\sigma}$ are used interchangeably to denote the same quantity. Figure~\ref{fig:just}(d) indicates that $\textbf{D} \perp \textbf{E}_0$ enables optical gain even in the low-scattering regime ($\gamma \ll \omega$), which was inaccessible in the setup of the Ref.~\cite{PhysRevLett.130.076901} due to their assumption that $\textbf{D} \parallel \textbf{E}_0$.

We find that TE-polarized light yields the highest optical gain at $\phi = 270^{\circ}$ and that the transmission peaks at $\phi = 90^{\circ}$ and $\phi = 270^{\circ}$ are not equal. Figure~\ref{fig:just}(c)-(d) shows the transmission at $\phi = 90^{\circ}$ (c) and $\phi = 270^{\circ}$ (d) while sweeping the tilt angle $\Psi$ and the ellipticity angle $\chi$ of the incident field [see Appendix E for further details concerning the relation between E-field components and Poincare sphere parameters]. Comparing Fig.~\ref{fig:just}(c) with Fig.~\ref{fig:just}(d), $T_{\textrm{max}}(\phi = 270^\circ) > 1$ whereas $T_{\textrm{max}}(\phi = 90^\circ) < 1$. The fact that $T_{\textrm{max}}(\phi = 90^\circ) < T_{\textrm{max}}(\phi = 270^\circ)$, as shown in  Fig.~\ref{fig:just}(c), is related to the sign of the power dissipated from the optical field $\textbf{E}_\omega$ to the material. From the non-Hermitian contribution, the dissipated power $p_{\textrm{dis}} = \textrm{Re}[\textbf{E}_{\omega}^{*}\cdot\boldsymbol{\sigma}^{\textrm{NH}}\cdot\textbf{E}_{\omega}]/2$ is positive for $T(\phi = 90^\circ)$ and negative for $T(\phi = 270^\circ)$, meaning optical loss and gain, respectively.

\begin{figure*}[t]
\includegraphics[width = 0.95\linewidth]{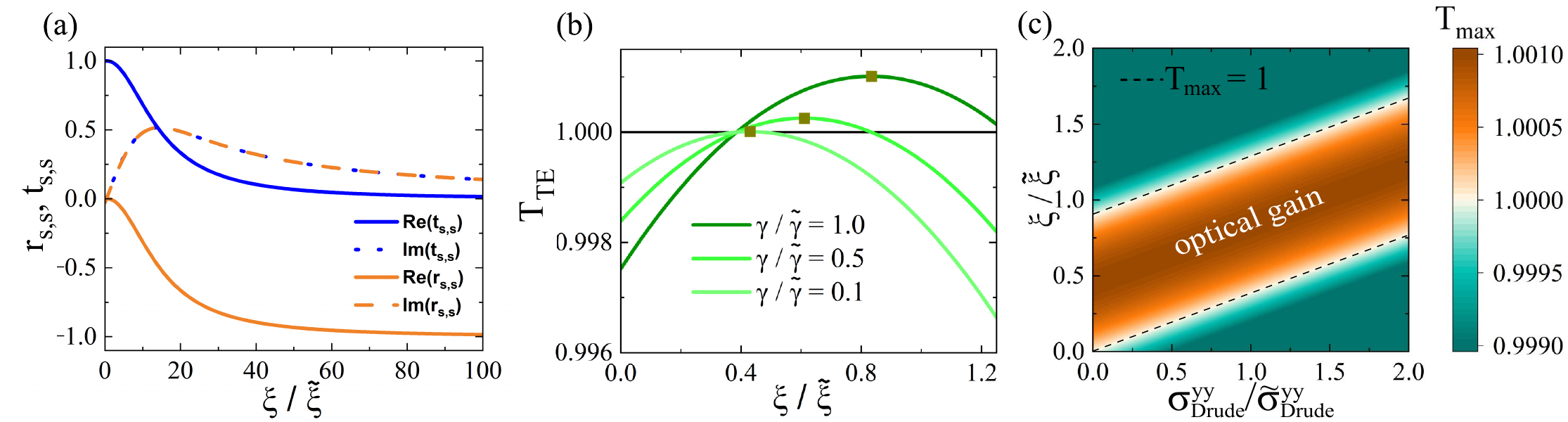}
\caption{Increasing $\xi$ does not guarantee more optical gain with only $\textbf{D}$ present. (a) Reflection coefficient of an TE-polarized wave as a function of $\xi$. (b) Transmittance of an TE-polarized wave as a function of $\xi$, where $\tilde{\gamma} = 10^{11}\textrm{rad/s}$. (c) Transmittance as a function of $\xi$ and $\sigma^{yy}_{\textrm{Drude}}$ at the optimal polarization; $\tilde{\sigma}^{yy}_{\textrm{Drude}}$ is fixed and given by Eq.~\ref{eq:drude}. In all panels, $\tilde{\xi} = 15.3$ THz; $\phi = 270 ^{\circ}$; and the values of $\rho$, $\omega$, $\epsilon_1$, and $\epsilon_2$ are the same as in Fig.~\ref{fig:just}.}
\label{fig:barry}
\end{figure*}

Physically, this phenomenon may be understood by analyzing the non-equilibrium current density $\textbf{J} = \sum_k f\textbf{v}$ induced by the simultaneous presence of optical and static fields. First, when $\textbf{E}_0$ points along $\pm\hat{\textbf{x}}$ it induces an anomalous velocity, proportional to $ \textbf{E}_0\times \boldsymbol{\Omega}$~\cite{RevModPhys.82.1959}, along the $\mp\hat{\textbf{y}}$ direction. Recall that we always assume $\hat{\textbf{y}}\parallel \textbf{E}_{\omega}$ for TE-polarized normally incident optical field. Second, assuming linear response to the optical fields, the induced non-equilibrium distribution function $f$ oscillates in-phase with $\textbf{E}_{\omega}$. Thus, the average power dissipated from the optical field to the anomalous current density, i.e., the contribution derived from the anomalous velocity, is 
\begin{equation}
    \langle \textbf{J}^{*}\cdot \textbf{E}_{\omega} \rangle \propto (\textbf{E}_\omega^*\cdot \textbf{D})\textbf{E}_0 \times \hat{\textbf{z}} \cdot \textbf{E}_{\omega},
\end{equation}
where the vector representing the non-vanishing components of the BCD is $\textbf{D} =  \sum_{n\textbf{k}}f^0_{n\textbf{k}}\nabla_{\textbf{k}}\Omega_{n\textbf{k}}^z$ can be chosen such that $\textbf{E}_\omega^*\cdot \textbf{D}$ is a positive quantity. When $\phi = 270^{\circ}$, i.e., the $\textbf{E}_0$ direction is $\hat{\textbf{x}}$, the anomalous velocity $\propto \textbf{E}_0 \times \hat{\textbf{z}}$ is antiparallel to $ \textbf{E}_{\omega}$, resulting in a negative dissipated power, i.e., optical gain. When $\phi = 90^{\circ}$, i.e., the $\textbf{E}_0$ direction is $-\hat{\textbf{x}}$, the anomalous velocity is parallel to $ \textbf{E}_{\omega}$, resulting in positive dissipated power, i.e., loss. This highlights not only the importance of the orientation of $\textbf{E}_0$ relative to $\textbf{D}$, but also the linear response character of the effect, that provides in-phase current-density oscillations with the optical field, without which the time-average vanishes. If $\textbf{E}_\omega^*\cdot \textbf{D}$ is a negative quantity and $\textbf{E}^{*}_\omega$ points along $\hat{\textbf{y}}$, $\textbf{D}$ points along $-\hat{\textbf{y}}$, and hence our conclusions hold the same with the prescription $\phi \rightarrow \phi + 180^{\circ}$.

In addition to enabling the condition for maximal transmission, our results show that at oblique incidence, the angle $\phi \not\in [90^{\circ}, 270^{\circ}]$ allows for the largest optical gain, which induces off-diagonal components in the effective conductivity $\boldsymbol{\sigma}_{\textrm{eff}}(\omega)$. This leads to elliptically polarized light that optimizes transmission, in agreement with Ref.~\cite{PhysRevLett.130.076901}. A slight increase in the optical gain is also observed. However, the impact of the incidence angle will not be addressed here. For the remainder of this section, we restrict our analysis to normally incident light with $\phi = 270^{\circ}$.

In contrast with previous studies~\cite{PhysRevLett.130.076901}, our results reveal that the optical gain in the configuration discussed above is bounded with respect to both the Berry curvature dipole (BCD) and the magnitude of the static bias. That is, the gain does not grow indefinitely with increasing $\xi$, but instead reaches a maximum before decreasing. This behavior arises from the dependence of the reflected wave on $\xi$, which was previously assumed to be negligible. Since $\xi \propto  E_0 D_0$, the enhancement of $\xi$ can be attributed to increases in either $D_0$ or $E_0$. As shown in Fig.\ref{fig:barry}(a), where we fix $\tilde{\xi} = 15.3$ THz, which corresponds to $\tilde{D}_0 =40$ nm and $\tilde{E}_0 = 8 \times 10^4$ V$\cdot$m$^{-1}$, and plot the reflection coefficient with respect to the dimensionless quantity $\xi / \tilde{\xi}$, the reflected field grows in amplitude and becomes increasingly out of phase with the incident wave as $\xi$ increases, causing the total field $\mathbf{E}_{\omega} = \mathbf{E}_\textrm{inc} + \mathbf{E}_\textrm{ref}$ to approach zero in the large-$\xi$ limit. This destructive interference limits the achievable gain. The non-monotonic dependence of gain on $\xi$ can also be understood from Eq.~(\ref{grand}) and the resonance conditions: increasing $\xi$ scales the effective conductivity, which is proportional to $1 / (\gamma - i\omega)$. Since both $\text{Re}(\sigma^{yy}_{\textrm{eff}})$ and $\text{Im}(\sigma^{yy}_{\textrm{eff}})$ scale together, increasing $\xi$ beyond the point where $\sigma^{yy}_{\textrm{eff}}$ most closely satisfies the resonance leads to a detuning effect and a corresponding reduction in gain.

The maximum gain can be enhanced by increasing the scattering rate, $\gamma$. While this trend is consistent with previous studies~\cite{PhysRevLett.130.076901}, the mechanism behind the enhancement differs in our setup. Earlier works attributed the gain increase to impedance mismatch arising from the growth of $\boldsymbol{\sigma}^\textrm{H}_\textrm{D}$ as $\gamma$ decreases. In contrast, for the configuration considered here, where $\textbf{D} \perp \textbf{E}_0$, the Hermitian EO contribution $\boldsymbol{\sigma}^\textrm{H}_\textrm{D}$ vanishes identically. Instead, we find that $\text{Im}(\sigma^{yy}_{\textrm{eff}})\propto \omega / (\gamma^2 + \omega^2)$, which tends to zero as $\gamma \rightarrow \infty$, enabling an increase in the attainable maximum gain when $\textbf{D} \perp \textbf{E}_0$. This behavior is illustrated in Fig.~\ref{fig:barry}(b), which shows the transmission of a TE-polarized wave a function of the normalized $\xi$ for different scattering rates $\gamma$, normalized by $\gamma_0 = 10^{11}\textrm{rad/s}$. The transmission peak is highlighted for each case with a symbol. The values of $\gamma$ are restricted to satisfy the dilute impurity limit, $\gamma \ll \omega$, specifically $\gamma \leq 1 \cdot 10^{11}~\textrm{rad/s}$.

Furthermore, our results show that the bounded nature of the gain remains robust in the presence of anisotropy. To demonstrate this, we allow for unequal components of the Drude conductivity, by varying $\sigma_{\textrm{Drude}}^{yy}$ and analyzing the optical gain phase space in Fig.~\ref{fig:barry}(c), where we fix $\tilde{\sigma}^{yy}_{\textrm{Drude}}$ to the conductivity value in Eq.~(\ref{eq:drude}). The results reveal that anisotropy can either facilitate or hinder the onset of optimal optical gain at lower values of $\xi$. Here, the bound between gain and no-gain regions in the phase space is a linear function of anisotropy for all $\xi$. The mechanism can be understood as being equivalent to scaling the prefactor $1/(\gamma - i\omega)$, effectively bringing $\sigma^{yy}_{\textrm{eff}}$ closer to satisfying the resonance condition. This counterintuitive outcome offers an unusual means of controlling the maximum gain. In addition to strain engineering, which can be used to achieve appropriate combinations of anisotropy and $\xi$ at fixed $E_0$, anisotropic van der Waals heterostructures composed of intrinsically anisotropic 2D materials have also been shown to support both sizable BCD and a certain degree of anisotropy\cite{PhysRevLett.133.146605, Lu2024}. Figure~\ref{fig:barry}(c) also shows that the independent tunability of the real and imaginary parts of $\boldsymbol{\sigma}_\textrm{eff}$ may aid in achieving a higher gain. This conclusion is reinforced by Fig.~\ref{fig:barry}(b), where $\text{Re}(\boldsymbol{\sigma}_{\textrm{eff}})$ is varied independently of $\text{Im}(\boldsymbol{\sigma}_{\textrm{eff}})$, apart from their magnitudes.

Next, we show that optical gain can also be attained independently through the magnetoelectric EO effect, although with distinct signatures.

\begin{figure*}[t]
\includegraphics[width = 0.90\linewidth]{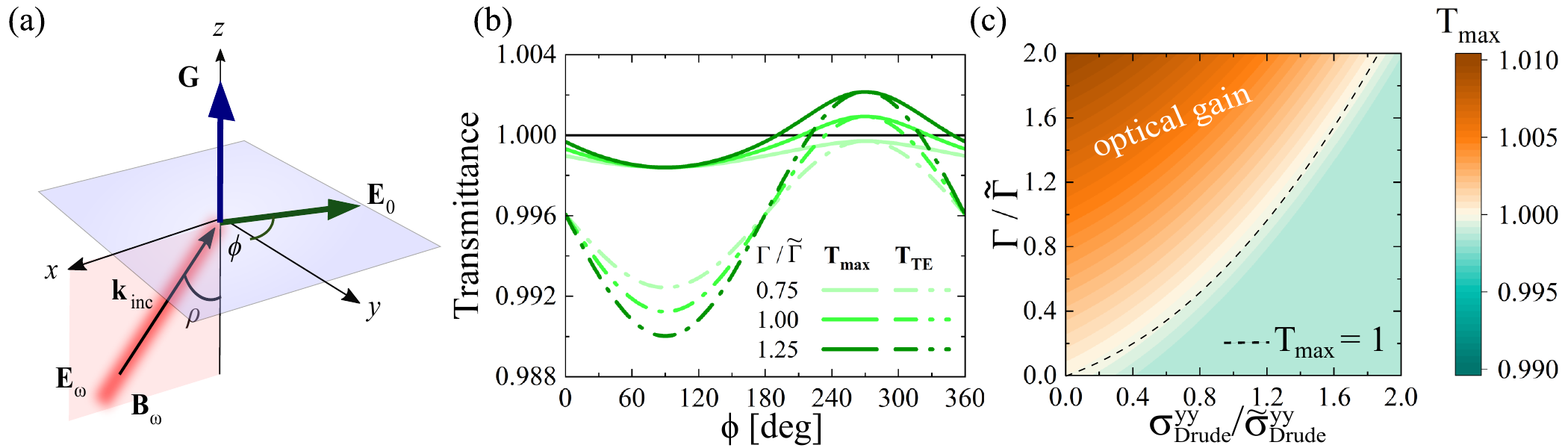}
\caption{Exploring how optical gain in the presence of only $\textbf{G}$ depends on $\phi$, $\Gamma$, $\sigma_{\textrm{Drude}}^{yy}$, and polarization. $\phi = 270^{\circ}$ and TE-polarized light once again maximize gain. (a) Schematic including only $\mathbf{G}$ and $\mathbf{E}_{\textrm{0}}$. The optical field is now incident at an angle $\rho \neq 0$ and the coordinate axis is oriented such that the optical field is incident in the $xz$-plane. (b) Transmittance of optimally- and TE-polarized light with respect to $\textbf{E}_0$ direction, with $G_0 = 200$ m$^2 \cdot$V$^{-1}\cdot$s$^{-1}$ and ${E}_{\textrm{0}} = 8 \times 10^4$ V$\cdot$m$^{-1}$. (c) Transmittance of optimally-polarized light as a function of $\sigma^{yy}_{\textrm{Drude}}$ and $\Gamma$, where $\tilde{\Gamma} = 0.053$ and $\phi = 270^{\circ}$. In panels (b) and (c), $\rho = 45^{\circ}$; the values of $\omega$, $\epsilon_1$, $\epsilon_2$, and $\tilde{\sigma}^{yy}_{\textrm{Drude}}$ are the same as in Fig.~\ref{fig:just}.}
\label{fig:magn}
\end{figure*}

\subsection{C. Optical Gain through magnetoelectric electro-optic effects}

Here, we focus on the optical gain induced solely by the magnetoelectric EO effect. As discussed previously, we assume that the 2D system supports exclusively out-of-plane components for $\boldsymbol{\Omega}_{n\textbf{k}}$ and $\textbf{m}_{n\textbf{k}}$, such that there is a single non-vanishing component for the EO MET $\textbf{G}$: $G_0 \equiv G^{zz} = \sum_{n\textbf{k}} (-\partial f^0_{n\textbf{k}}/\partial \epsilon_{n\textbf{k}}) \Omega^z_{n\textbf{k}} m^z_{n\textbf{k}}$. Furthermore, to study the effect of $G_0$ alone, we assume that the BCD vanishes. This is made possible by certain point group symmetries~\cite{dichroism}, and will be taken here as an implicit assumption. In the following, we take $\Gamma = \epsilon_2 E_0G_0 /c$ to represent the magnitude of the bias-induced magnetoelectric coupling and focus on the general features of the gain induced by $\boldsymbol{\sigma}_{\textrm{G}}$.

Due to $\mathbf{G}$ having only a $G^{zz}$ component, orienting the $xy$-axes is an arbitrary choice. We chose this orientation such that the light is incident in the $xz$-plane, i.e. the direction of incident light propagation \textit{fixes} the $xy$-axis orientation. From here, we may use Eq.~\ref{eqsigma_G} together with the scattering formalism developed above. This treatment is in stark contrast to the discussion of the optical gain due to the BCD, since there we assumed that the $xy$-axis orientations were dependent on the orientation of $\mathbf{D}$ in the material. This distinction, although important to understand, still allows for us analyzing the combined effect of the MET and BCD (as done in the next section).  

Figure~\ref{fig:magn}(b) shows the transmission as a function of the in-plane orientation angle $\phi$ of $\textbf{E}_0$, revealing that the optimal configuration occurs at $\phi = 270^{\circ}$, akin to the BCD scenario. However, it is important to note that this configuration now refers to the relative alignment between $\textbf{E}_0$ and the incidence plane $xz$. We emphasize that the shape of the $\boldsymbol{\sigma}_{\textrm{G}}$ tensor, given in Eq.~(\ref{eqsigma_G}), relies on our convention for the place of incidence of $\textbf{E}_\omega$ as being the $xz$ plane. If the plane of incidence was chosen to be the $yz$-plane, one would find $\sigma_\textrm{G}^{xx}, \sigma_\textrm{G}^{yx} \neq 0$ and $\sigma_\textrm{G}^{xy}, \sigma_\textrm{G}^{yy} = 0$, which is consistent with $\textbf{G}$ only coupling with $B^z$ [see Appendix C for further details]. 

We also note that, similar to the BCD case, TE-polarized light maximizes transmission at $\phi = 270^{\circ}$, which is simply explained by the coupling between $B^z$ and $\textbf{E}_0$ via $\textbf{G}$. Thus, for the remainder of this section, we limit our discussion to TE-polarized light traveling toward the system at oblique incidence. Due to the similarities between the $\boldsymbol{\sigma}_{\textrm{G}}(\omega)$ and $\boldsymbol{\sigma}_{\textrm{BCD}}(\omega)$ tensor shapes for the $\phi = 270^{\circ}$ and $xz$ incidence plane setup, we proceed by comparing the response signatures induced by the BCD and the bias-induced magnetoelectric coupling.  

Unlike the gain induced by the BCD discussed previously, the optical gain resulting from the $\boldsymbol{\sigma}_{\textrm{G}}(\omega)$ tensor increases monotonically with $\Gamma$ for $0 \leq \Gamma/\tilde{\Gamma} \leq 2$, as shown in the effective conductivity anisotropy versus $\Gamma$ phase space displayed in Fig.~\ref{fig:magn}(c). Here, $\tilde{\Gamma}$ denotes a reference value of the magnetoelectric coupling, corresponding to an applied in-plane electric field $\tilde{E}_0 = 8 \times 10^4$ V/m and $G_0 = 200$ m$^2 \cdot \textrm{V}^{-1} \cdot \textrm{s}^{-1}$. The monotonic enhancement of optical gain with $\Gamma$, or $E_0$, arises from the dissipative (real) surface-current responses in $\boldsymbol{\sigma}_\textrm{G}(\omega)$. Although both the BCD and bias-induced magnetoelectric EO effect lead to a Drude-like perturbation of the distribution function $\delta f \propto 1/(\gamma - i\omega)$ (deriving from the balance between relaxation effects and the optical field tendency to drive the system out-of-equilibrium), the coupling between the magnetic moment of Bloch electrons and the $\textbf{B}_{\omega}$ optical field also induces a Zeeman-like shift in the equilibrium distribution $f^0_{n\textbf{k}}$, which grows with $\omega$ in a out-of-phase manner with respect to $\textbf{E}_{\omega}$~\cite{PhysRevB.110.115421}. This effect leads to $|\text{Im}(\sigma_\textrm{G}^{yy})| \ll |\text{Re}(\sigma_\textrm{G}^{yy})|$ when $\gamma \ll \omega$, which allows for tuning $\sigma^{yy}_{\textbf{eff}}$, where $\boldsymbol{\sigma}_{\textrm{eff}}(\omega) = \boldsymbol{\sigma}_{\textrm{Drude}}(\omega) + \boldsymbol{\sigma}_{\textrm{G}}(\omega)$ , closer to satisfying TE-mode resonance condition $2\cos\rho + z_0\text{Re}(\sigma^{yy}_{\textrm{eff}}) = 0$ without increasing the imaginary component. This is in contrast with the BCD scenario, where increasing $\xi$ would bring $\sigma^{yy}_{\textrm{eff}}$ closer to satisfying condition $2\cos\rho + z_0\text{Re}(\sigma^{yy}_{\textrm{eff}}) = 0$ but also to deviate from the second TE-mode resonance condition $\text{Im}(\sigma^{yy}_{\textrm{eff}}) = 0$. The contrast may be seen by comparing Fig.~\ref{fig:barry}(c) with Fig.~\ref{fig:magn}(c). Optical gain in the latter case is best achieved when $\sigma^{yy}_\textrm{Drude} / \tilde{\sigma}^{yy}_{\textrm{Drude}} = 0$, since that is when $\text{Im}(\sigma^{yy}_{\textrm{eff}})$ is closest to $0$. However, the sole presence of $\textbf{G}$ does not allow us to fully access the resonance either. Even if engineering $\sigma^{yy}_\textrm{Drude} / \tilde{\sigma}^{yy}_{\textrm{Drude}} = 0$ were possible, a finite $\gamma$ prevents the second resonance condition $\text{Im}(\sigma^{yy}_{\textrm{eff}}) = 0$ from being achieved. 

This motivates a closer examination of the combined effects of $\boldsymbol{\sigma}_{\textrm{G}}(\omega)$ and $\boldsymbol{\sigma}_{\textrm{BCD}}(\omega)$ on the optical gain, which we explore in the following section.

\subsection{D. Optical Gain in the General Case}

\begin{figure*}[t]
\includegraphics[width = 0.85\linewidth]{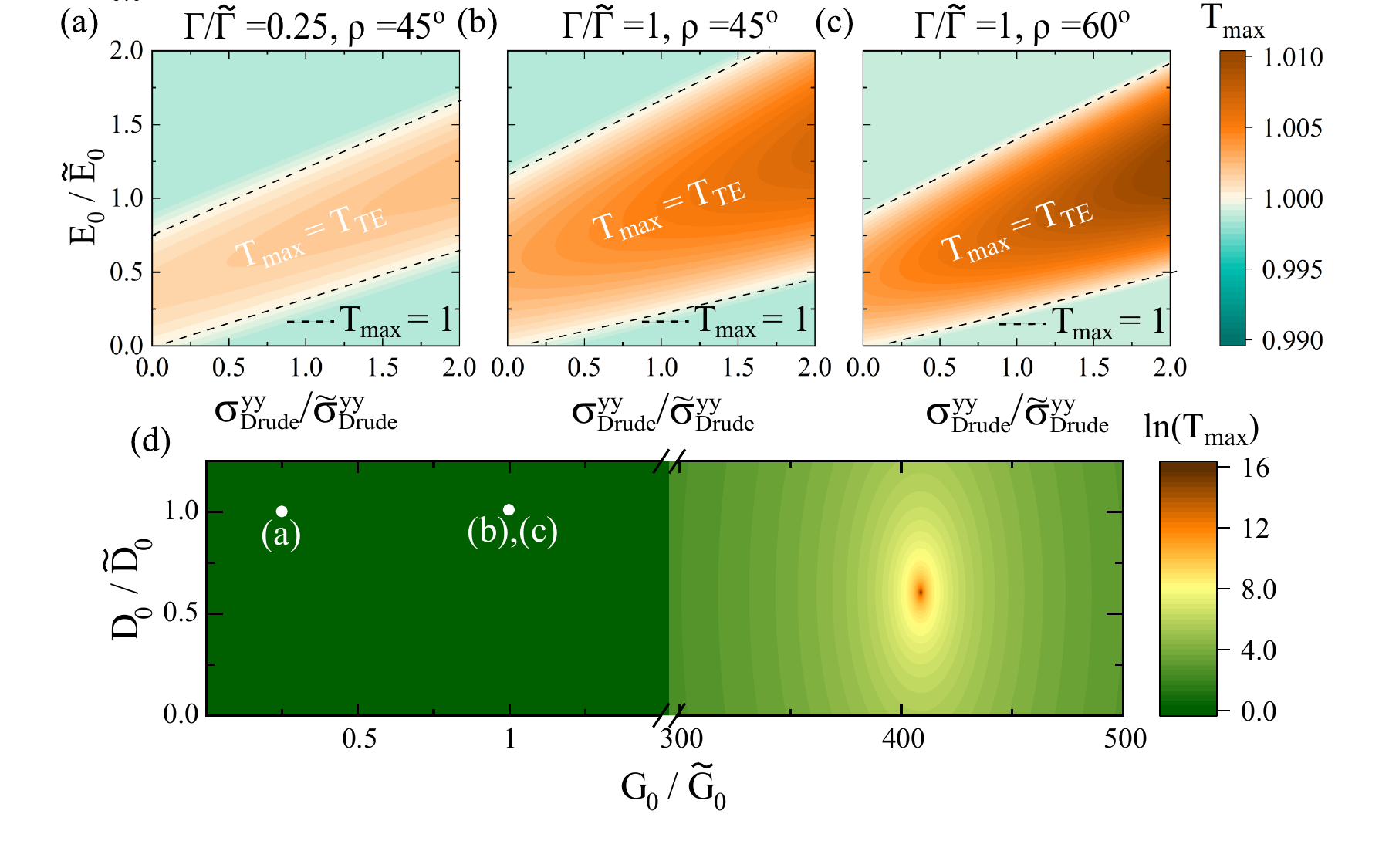}
\caption{TE-resonance condition may be accessed by tuning $E_0$, $D_0$, and $G_0$. (a)-(c) Transmittance as of optimally-polarized light as a function of ${\sigma}^{yy}_{\textrm{Drude}}$ and in-plane bias magnitude $E_0$, with $\tilde{E}_{\textrm{0}} = 8 \times 10^4$ V$\cdot$m$^{-1}$ and $D_0 = 40$ nm. The value of $\tilde{\sigma}^{yy}_{\textrm{Drude}}$ is the same as in Fig.~\ref{fig:just}. (d) Natural log of transmittance as a function of $D_0$ and $G_0$, where $\tilde{D}_0 = 40$ nm, $\tilde{G}_0 = 200$ m$^2 \cdot$V$^{-1}\cdot$s$^{-1}$, $E_0 = 8 \times 10^4$ V$\cdot$m$^{-1}$, $\rho = 45^{\circ}$, and ${\sigma}^{yy}_{\textrm{Drude}}/\tilde{\sigma}^{yy}_{\textrm{Drude}}=1$. The points which correspond to the values of $D_0$ and $G_0$ used in panels (a)-(c) are marked in white in panel (d). In all panels, $\phi = 270^{\circ}$ and $\theta = 0^{\circ}$. Values of $\omega$, $\epsilon_1$, and $\epsilon_2$ are the same as in Fig.~\ref{fig:just}.}
\label{fig:comb}
\end{figure*}

We discuss here the features of optical gain when both $\boldsymbol{\sigma}_{\textrm{G}}(\omega)$ and $\boldsymbol{\sigma}_{\textrm{BCD}}(\omega)$ are simultaneously non-vanishing. The total effective conductivity becomes $\boldsymbol{\sigma}_{\textrm{eff}}(\omega) = \boldsymbol{\sigma}_{\textrm{Drude}}(\omega) + \boldsymbol{\sigma}_{\textrm{BCD}}(\omega) + \boldsymbol{\sigma}_{\textrm{G}}(\omega)$. Since the optical field only couples to the $\boldsymbol{\sigma}_{\textrm{G}}(\omega)$ at oblique-incidence, we only consider $\rho \neq 0$ henceforth. We also only consider the configuration with $\theta = 0$ and $\phi = 270^{\circ}$. These parameters were verified to yield maximum transmission in the general case, as discussed in previous sections. Moreover, this configuration offers a clean scenario in which only $\sigma^{yy}_{\textbf{eff}}(\omega)$ is modified, making it more suitable for isolating and analyzing the contributions of the two effects.

We find that optical gain may be enhanced by the presence of sufficiently large $G_0$, implying that we are able to come closer to satisfying the resonance conditions using it, as seen in Fig.~\ref{fig:comb}(d). In Figs.~\ref{fig:comb}(a)-(c), the trend in the gain as the anisotropy is increased reflects that of the Berry dipole, where the region of maximum transmission is given by a linear relation between $\textbf{E}_0$ and $\sigma^{yy}_\textrm{Drude}$, as can be observed in Fig.~\ref{fig:barry}(b). This implies that $\text{Im}(\sigma^{yy}_\textrm{eff})$ is once again the limiting factor in satisfying both resonance conditions. When $\text{Im}(\sigma^{yy})$ is close to $0$, we observe that either increasing $G_0$ or coupling to it more effectively by increasing the angle of incidence leads to more gain. This can be seen when comparing Fig.~\ref{fig:comb}(a) with Fig.~\ref{fig:comb}(b), and Fig.~\ref{fig:comb}(b) with Fig.~\ref{fig:comb}(c), respectively. We emphasize the role of the TE-polarization due to its coupling to $\textbf{E}_0$ via $\textbf{G}$ (optical gain region in panels (a)-(c) of Fig.~\ref{fig:comb} correspond incident light which is TE-polarized).  

Tunable $\textbf{D}$ and $\textbf{G}$ simultaneously present allow for satisfying both resonance conditions. Indeed, 

\begin{equation}
    2\cos{\rho} + z_0\sigma^{yy} = \frac{2\cos\rho(\gamma - i\omega) + (\alpha' + \alpha(D_0) + \beta(G_0)i\omega)}{\gamma - i\omega},
    \label{zero}
\end{equation}
where $\alpha'$ captures the Drude contribution; $\alpha(D_0)$ the Berry dipole effects from Eqs.~(\ref{eq:Hermitian}) and (\ref{eq:bnh}); and $\beta(G_0)$ captures the effect of the magnetic moment. Clearly a zero of Eq.~(\ref{zero}) exists for a suitable choice of $\alpha'$, $\alpha$, and $\beta$. This zero is illustrated in Fig.~\ref{fig:comb}(d), where we show the TE-resonance and the points ($G_0$,$D_0$) which correspond to the values used in Figs.~\ref{fig:comb}(a)-(c). Drawing a parallel between this system and a damped harmonic oscillator driven near its natural frequency, tuning $D_0$ and $G_0$ such that condition 2) (Im$(\sigma^{yy})$ = 0) is met is akin to when the harmonic oscillator's damping vanishes. Achieving this, however, would require a suitably larger $G_0$, which hasn't been reported yet. Promising avenues to enable such larger $G_0$ values may include Moir\'e quantum matter, which has been demonstrated to enable giant orbital moments and Berry curvatures of Bloch electrons and gigantic Berry curvature dipoles~\cite{goldhaber, Pantaleon2021}.

\section{IV. Conclusions}

We have shown that a resonant (scattering) TE mode may be excited to yield optical gain in biased 2D metals which possess a BCD and non-zero MET. To these ends, we developed the system's conductivity using the semi-classical Boltzmann transport theory and found that the TE mode is accessed when the anomalous electron velocity is perpendicular to the bias. The resonance occurs between the anomalous velocity imparted on the electrons and the optical field. Within the limit linear response of the current density to the optical field, we have shown that this resonance, with suitable tuning of the system, may be engineered to effectively be lossless. This finding provides a promising avenue for realizing optical gain using intraband processes and the unique momentum textures of Bloch electrons on the Fermi surface. Further studies may take one of two directions: one direction which discovers materials which possess the large BCD and MET necessary to fully access the resonant TE mode and another direction which analyzes how bound modes interact with the system. The latter direction is promising, due to the confined nature of plasmon interactions, which may allow for large plasmonic gain without the need for an unrealistically large BCD and MET.


\section{Acknowledgments} 
The authors acknowledge partial support from Office of Naval Research MURI grant N00014-23-1-2567. LMM acknowledges projects PID2023-148359NB-C21 and CEX2023-001286-S (financed by MICIU/AEI/10.13039/501100011033) and the Government of Aragon through Project Q-MAD. NRL acknowledges support from the University of Minnesota's Pathways to Graduate School: Summer Research Program.

\section{Appendix A: Explicit Calculation of the Conductivity Coefficients for a Sample toy model}\label{app}
Here, we derive practical analytical expressions for the various conductivity coefficients in a 2D gapped Dirac electron system. These results, primarily used as references in the main text, are presented explicitly here for completeness. The Hamiltonian is $H_{\eta} = \hbar v_F (\eta\sigma_x k_x + \sigma_y k_y) + (\Delta/2)\sigma_z$, where $\eta = \pm1$ is the valley index, $v_F$ is the fermi velocity, $\Delta$ quantifies the size of the energy gap and $\sigma_j$, $j=x,y,z$, are the Pauli matrices. Because the BCD for such a toy model has already been calculated elsewhere~\cite{Sinha2022}, we are focusing on the derivation of the Drude conductivity and the linear magnetoelectric electro-optical effect coefficient. In the derivation that follows, we assume the Fermi level satisfies $\mu > \Delta/2$, ensuring that the conduction band remains populated and the system exhibits metallic behavior. Analogous expressions can be obtained for $\mu < -\Delta/2$, in which case a finite hole population governs the metallic response.  

The orbital contribution to the magnetic moment of Bloch wave packets is~\cite{RevModPhys.82.1959}
\begin{equation}
    m_\eta^z (\mathbf{k}) = \frac{\eta e}{4\hbar}(\hbar v_F)^2 \frac{\Delta}{\left[\left(\frac{\Delta}{2}\right)^2 + (\hbar v_F k)^2\right]},
    \label{eq:orbmag}
\end{equation}
and the Berry curvature is given by
\begin{equation}
    \Omega_\eta^z(\mathbf{k}) = \frac{\eta}{4}(\hbar v_F)^2 \frac{\Delta}{\left[\left(\frac{\Delta}{2}\right)^2 + (\hbar v_F k)^2\right]^{1/2}},
    \label{eq:berry}
\end{equation}
where $\eta = \pm 1$, depending on the valley. The conductivities $\sigma_E (\omega)$ and $\sigma_B(\omega)$ are equal to
\begin{align}
    \sigma_E(\omega) &= \frac{e^2\tau}{1 - i\omega \tau}\int \frac{d^2k}{(2\pi)^2}\left[-\frac{\partial f(\epsilon - \mu)}{\partial \epsilon_{\eta, k}} \right] v_{\eta, k}v_{\eta, k}, \\
    \sigma_B(\omega) &=  e\frac{i\omega\tau}{i \omega \tau - 1} \int \frac{d^2k}{(2\pi)^2}\left[-\frac{\partial f(\epsilon - \mu)}{\partial \epsilon_{\eta, k}} \right] m_\eta^z(\mathbf{k}) \Omega_\eta^z(\mathbf{k}),
    \label{eq:sigma}
\end{align}
where $v_{\eta, k} = 1/\hbar \partial \epsilon_{\eta}(\mathbf{k})/\partial k$ is the velocity operator. The monolayer graphene bands are given by
\begin{equation}
    \epsilon_{\eta}(\mathbf{k}) = \pm \sqrt{\left(\frac{\Delta}{2}\right)^2 + (\hbar v_F k)^2},
    \label{eq:bands}
\end{equation}
so the velocity operator is thus
\begin{align}
    v_{\eta}(k) = \pm \left[ \left(\frac{\Delta}{2}\right)^2 + (\hbar v_F k)^2\right]^{-1/2} \hbar^2 v_F^2 k.
    \label{eq:vel}
\end{align}

Next, we explicitly derive the conductivities. We begin by addressing the Drude conductivity. 

First, we start changing the integration from $k-$space to integrating over the energies $\epsilon$. From Eq.~(\ref{eq:bands}), we obtain
\begin{equation}
    k = \pm \frac{1}{\hbar v_F}\sqrt{\epsilon^2 - \left(\frac{\Delta}{2}\right)^2},
    \label{eq:k}
\end{equation}
from which,
\begin{equation}
    dk = \frac{1}{\hbar v_F} \frac{\epsilon}{\left[ \epsilon^2 - \left(\frac{\Delta}{2}\right)^2 \right]^{1/2}}d\epsilon.
    \label{eq:dk}
\end{equation}
and
\begin{equation}
    k dk = \frac{1}{(\hbar v_F)^2}\epsilon d\epsilon.
\end{equation}
Therefore, 
\begin{equation}
\int  F(k) \frac{d^2 k}{(2\pi)^2} = \frac{1}{2\pi} \int_0^\infty \frac{\epsilon}{(\hbar v_F)^2} F(\epsilon) d\epsilon.
\end{equation}
Using Eqs.~(\ref{eq:k}) and (\ref{eq:vel}),
\begin{align}
    v_{\eta}(\epsilon) = \frac{v_F}{\epsilon}\sqrt{\epsilon^2 - \left(\frac{\Delta}{2}\right)^2},    
\end{align}
where we have assumed $d^2 k = k dk d\phi$. replacing this result into 
Eq.~(\ref{eq:sigma}), it follows that
\begin{align}
    \sigma_E &\propto \frac{1}{2\pi} \int_0^\infty \left[ - \frac{\partial f (\epsilon-\mu)}{\partial \epsilon}\right] v^2(k) k dk \notag \\
    &\sim \frac{1}{2\pi}\int_0^\infty \delta (\epsilon - \mu) \frac{v_F^2}{\epsilon^2}\left[ \epsilon^2 - \left(\frac{\Delta}{2}\right)^2\right] \frac{1}{(\hbar v_F)^2}\epsilon d\epsilon \notag \\
    &\sim \frac{1}{2\pi \hbar^2}\int_0^\infty \delta(\epsilon-\mu) \left[ \epsilon^2 - \left(\frac{\Delta}{2}\right)^2\right]\frac{d\epsilon}{\epsilon}\notag \\
    &\sim \frac{1}{2\pi\hbar^2} \left[ \mu^2 - \left(\frac{\Delta}{2}\right)^2 \right] \frac{1}{\mu} \notag \\
    &\sim \frac{\mu}{2\pi \hbar^2}\left[ 1 - \left( \frac{\Delta}{2\mu} \right)^2 \right],
\end{align}
where we have assumed that the derivative of the Fermi-Dirac distribution with respect to the energy is approximately a Dirac delta function in the limit $T\rightarrow 0$. Including the pre-factor $\tau/(1 - i \omega \tau)$, the conductivity becomes
\begin{equation}
    \sigma_E(\omega) = \frac{e^2 \tau}{1 - i \omega \tau} \frac{\mu}{2\pi \hbar^2}\left[1 - \left(\frac{\Delta}{2\mu}\right)^2\right],
\end{equation}
valid for $\mu > \Delta/2$. 

Next, we address linear magnetoelectric electro-optic effect coefficient calculation. Following along the same lines, we start by changing Eqs.~(\ref{eq:orbmag}) and (\ref{eq:berry}) from momentum to energy:
\begin{align}
    \Omega_\eta^z (\mathbf{k}) &= \frac{\eta}{4}(\hbar v_F)^2 \frac{\Delta}{\epsilon^3} \\
    m_\eta^z (\mathbf{k}) &= \frac{\eta e}{4\hbar}(\hbar v_F)^2 \frac{\Delta}{\epsilon^2}.
\end{align}
Hence,
\begin{align}
    \sigma_B(\omega) &\propto \int \frac{d^2 k}{(2\pi)^2}\left[ -\frac{f (\epsilon - \mu)}{\partial \epsilon}\right]\Omega_\eta^z (\mathbf{k}) m_\eta^z (\mathbf{k}) \notag \\
    &\sim \frac{1}{2\pi} \int_0^\infty \delta(\epsilon - \mu) \frac{e(\hbar v_F)^2}{16 \hbar}\frac{\Delta^2}{\epsilon^5}\epsilon d\epsilon \notag \\
    &\sim \frac{e(\hbar v_F)^2 \Delta^2}{32\pi\hbar \mu^4}.
\end{align}
And finally,
\begin{equation}
    \sigma_B(\omega) = \frac{e^2}{\hbar}\frac{i\omega\tau}{i \omega \tau - 1} \frac{(\hbar v_F)^2 \Delta^2}{32\pi \mu^4},
\end{equation}
valid for $\mu > \Delta/2$. Comparing with the expression utilized in the main text, we identify $\Gamma = (\hbar v_F \Delta)^2/32 \pi \mu^4$.

\section{Appendix B: Berry dipole conductivity tensor for the 2D case}\label{appB}

Following Ref.~\cite{PhysRevB.110.115421}, the Hermitian (H) and non-Hermitian (NH) conductivity tensors originating from the BCD (BCD) are 
\begin{eqnarray}
\boldsymbol{\sigma}_{\textrm{D}}^{\textrm{H}} = \displaystyle -\frac{e^3\tau}{\hbar^2} \textbf{D} \cdot \textbf{E}_0,
   \label{eq42}
\end{eqnarray}
\begin{eqnarray}
\boldsymbol{\sigma}_{\textrm{D}}^{\textrm{NH}} = \displaystyle \frac{e^3\tau}{\hbar^2} \frac{1}{1-i\omega\tau} \textbf{F}_0 \cdot \textbf{D},
    \label{eq47}
\end{eqnarray}
respectively, where
\begin{eqnarray}
\textbf{F}_0 &= \left[
\begin{tabular}{ccc}
    $0$ &  $0$  & $E_0^y$ \\
    $0$ &  $0$  & $-E_0^x$ \\
    $-E_0^y$ & $E_0^x$ & $0$
\end{tabular}\right], \label{eq:F0} \\
\textbf{E}_0 &= \left[
\begin{tabular}{c}
    $E_0^x$ \\
    $E_0^y$  \\
    $0$ 
\end{tabular}\right], \label{eq:E0} \\
\textbf{D} &= \left[
\begin{tabular}{ccc}
    $0$ &  $0$  & $0$ \\
    $0$ &  $0$  & $0$ \\
    $D^x$ & $D^y$ & $0$
\end{tabular}\right],
\label{eq43}
\end{eqnarray}
where $D^{\alpha} = -\sum_{n\textbf{k}}f_{n\textbf{k}}^0 \nabla_{\textbf{k}}^{\alpha} \Omega^z_{n\textbf{k}}$ is the $\alpha = x,y$ component of the BCD. We assume that this Berry dipole derives from a single Berry curvature component $\Omega^z_{n\textbf{k}}$ in 2D systems. Thus, a trivial calculation renders 
\begin{eqnarray}
\boldsymbol{\sigma}_{\textrm{D}}^{\textrm{H}} = \displaystyle -\frac{e^3\tau}{\hbar^2}  \left[
\begin{tabular}{ccc}
    $0$ &  $-D^x E_0^x-D^y E_0^y$  & $0$ \\
    $D^x E_0^x+D^y E_0^y$&  $0$  & $0$ \\
    $0$ & $0$ & $0$
\end{tabular}\right],
    \label{eq44}
\end{eqnarray}
\begin{eqnarray}
\boldsymbol{\sigma}_{\textrm{D}}^{\textrm{NH}} = \displaystyle \frac{e^3\tau}{\hbar^2}  \frac{1}{1-i\omega\tau} \left[
\begin{tabular}{ccc}
    $D^x E_0^y$ &  $D^y E_0^y$  & $0$ \\
    $-D^x E_0^x$ &  $-D^y E_0^x$  & $0$ \\
    $0$ & $0$ & $0$
\end{tabular}\right].
    \label{eq45}
\end{eqnarray}
Next, we assume $D^x = -D_0 \sin\theta$, $D^y = D_0 \cos\theta$ and $E_0^x = -E_0 \sin\phi$, $E_0^y = E_0 \cos\phi$. We also note that it is possible to reduce the dimensionality of the conductivity tensors by taking the basis $\textbf{E}_{\omega} = E_{\omega}^x \hat{\textbf{x}} +  E_{\omega}^y \hat{\textbf{y}}$. Hence, 

\begin{eqnarray}
\boldsymbol{\sigma}_{\textrm{D}}^{\textrm{H}} = \displaystyle -\frac{e^3\tau}{\hbar^2}  D_0E_0 \left[
\begin{tabular}{cc}
    $0$ &  $-\cos(\theta - \phi)$ \\
    $\cos(\theta - \phi)$ &  $0$   \\
\end{tabular}\right],
    \label{eq53}
\end{eqnarray}
\begin{eqnarray}
\boldsymbol{\sigma}_{\textrm{D}}^{\textrm{NH}} = \displaystyle \frac{e^3\tau}{\hbar^2}  \frac{D_0E_0}{1-i\omega\tau} \left[
\begin{tabular}{ccc}
    $-\sin\theta \cos\phi$ &  $\cos\theta \cos\phi$ \\
    $-\sin\theta\sin\phi$ &  $\cos\theta\sin\phi$ \\
\end{tabular}\right].
    \label{eq54}
\end{eqnarray}

\section{Appendix C. magnetoelectric conductivity tensor for the 2D case}

Following Ref.~\cite{PhysRevB.110.115421}, the non-Hermitian (NH) conductivity tensors originating from the out-of-plane magnetic moment is 

\begin{equation}
    \boldsymbol{\sigma}_\textrm{G,B}^{\textrm{NH}} = \frac{e^2}{\hbar}\frac{i\omega\tau}{i\omega\tau -1} \mathbf{F}_0 \cdot \mathbf{G},
    \label{eq:G0B}
\end{equation}

where, $\textbf{F}_0$ and $\textbf{E}_0$ are defined the same as in Eqs.~\ref{eq:F0} and \ref{eq:E0} respectively; and 

\begin{equation} 
    \mathbf{G} = 
    \begin{bmatrix}
        0 & 0 & 0 \\
        0 & 0 & 0 \\
        0 & 0 & G_0
    \end{bmatrix}.
    \label{eq:G0}
\end{equation}

Equation~\ref{eq:G0} may be derived from $\textbf{G} = \sum_{n\textbf{k}}(-\partial f^0_{n\textbf{k}}/\partial \epsilon_{n\textbf{k}})\textbf{G}_{n\textbf{k}}$, with components related to $G^{\alpha\beta}_{n\textbf{k}} = \Omega^{\alpha}_{n\textbf{k}}m^{\beta}_{n\textbf{k}}$.

The extra subscript 'B' in Eq.~\ref{eq:G0B} is meant to indicate that the tensor is in the $B$-field basis, i.e. $\textbf{J}_\textrm{G} = \boldsymbol{\sigma}_\textrm{G,B}^{\textrm{NH}} \cdot \textbf{B}_{\omega}$. The remainder of this appendix is dedicated to transforming Eq.~\ref{eq:G0B} into the $E$-field basis. For a monochromatic incident wave, such as the one assumed in this paper, the B-field may be related to the E-field in the following way \cite{griffiths}:

\begin{equation}
    \begin{bmatrix}
        B^{x} \\
        B^{y} \\
        B^{z} 
    \end{bmatrix} =
    \frac{\sqrt{\epsilon_{r,i}}}{c}
    \begin{bmatrix}
        0 & -\hat{k}_{z,i} & \hat{k}_{y,i} \\
        \hat{k}_{z,i} & 0 & -\hat{k}_{x,i} \\
        -\hat{k}_{y,i} & \hat{k}_{x,i} & 0
    \end{bmatrix}
    \begin{bmatrix}
        E^{x} \\
        E^{y} \\
        E^{z} 
    \end{bmatrix}, 
    \label{basis}
\end{equation}

where $\hat{k}_{n,m}$ are the $n$th component of the $\hat{k}$ unit vector in the $i$th medium, and $c$ is the vacuum speed of light. $B^{z}$ is then equivalent to $\hat{-k_y}E^{x} + \hat{k_x}E^{y}$. Note that in the main text, we assume that the wave propagates in the $xz$-plane, i.e. $\hat{k}_{y,i} = 0 \Rightarrow \hat{k}_{x,i} = \sin{\rho_i}$. Since the quantity $\sqrt{\epsilon_r}\sin{\rho}$ is conserved by Snell's law, it suffices to only consider $\sqrt{\epsilon_{r,2}}\sin{\rho_{2}}$. Hence, rewriting Eq.~\ref{eq:G0B} in the $E$-field basis by using Eq.~\ref{basis} and combining the expression for $B$-field coupling with the known $E$-field coupling component allows us to re-write the conductivity in terms of the $E$-field. Note that the time-reversal breaking of the magnetic field is preserved under this operation, as the direction of the wave's propagation is captured by the $k_x$ and $k_y$ terms, which get absorbed into the conductivity expression. Thus,

\begin{eqnarray}    
    \boldsymbol{\sigma}_{\textrm{G}}^{\textrm{NH}} = \frac{e^2}{\hbar}\frac{i\omega\Gamma}{\gamma - i\omega}
    \begin{bmatrix}
        0 & -\sin\rho_2\cos\phi \\
        0 & -\sin\rho_2\sin\phi
    \end{bmatrix},
    \label{eq:magnet}
\end{eqnarray}

where $\Gamma = \epsilon_2 E_0G_0 /c$.

\section{Appendix D. transmission-maximizing polarization}

In the $E_x$, $E_y$ basis, the transmittance matrix is $\mathbf{T} = (n_2 \cos{\rho_{2}}/n_1 \cos{\rho_{1}})\mathbf{t}^{\dagger}\mathbf{t}$. This matrix is then $2 \times 2$ Hermitian \cite{PhysRevLett.130.076901}. The normalized column-vector $v$, which maximizes the expression $v^{\dagger}\mathbf{T}v$ is the eigenvector of $\mathbf{T}$ which corresponds to the maximum eigenvalue \cite{PhysRevLett.130.076901}.

We next show that this same treatment may be applied to find the optimal polarization in the $s$- and $p$-polarized basis. The magnitude of an arbitrary E-field is $\sqrt{E^{\dagger}E}$. This is then $\sqrt{\abs{E_x}^2 + \abs{E_y}^2 + \abs{E_z}^2}$. Plugging in values from Eq.~\ref{b1} in the main text, the magnitude of the E-field is then $\sqrt{\abs{a_s}^2 + \abs{a_p}^2}$. Thus, since the magnitude of the vector with TE-polarized and $p$-polarized components is equivalent to the magnitude of the E-field vector, we may maximize the transmittance with the procedure described above.

\section{Appendix E. General polarization using tilt and ellipticity angles}

Light may be described by four parameters known as the Stokes parameters \cite{Han1997}. Namely, they are \cite{Ren2015}: 

\begin{eqnarray}
    S_0 = I\\
    S_1 = Ip\cos{2\Psi}\cos{2\chi}\\
    S_2 = Ip\sin{2\Psi}\cos{2\chi}\\
    S_3 = Ip\sin{2\chi},\\
\end{eqnarray}

where $I$ is the intensity, $p$ is the degree of polarization, $\Psi$ is the tilt angle measured with respect to the +$x$-axis, and $\chi$ is the ellipticity angle. In the main text, we consider light which has intensity $I = 1$. We also limit our discussion to fully polarized light, i.e. $p = 1$. With these assumptions, we may relate the Stokes parameters to the $E_x$, $E_y$ components of the oscillating optical field using \cite{Han1997}:

\begin{eqnarray}
    S_1 = |E_x|^2 - |E_y|^2\\
    \label{eq:hopf1}
    S_2 = 2\text{Re}(E_xE_y^{*})\\
    \label{eq:hopf2}
    S_3 = 2\text{Im}(E_xE_y^{*}).\\
    \label{eq:hopf3}
\end{eqnarray}

The normalized expression for $E_x$, $E_y$, which satisfies Eqs.~\ref{eq:hopf1}-\ref{eq:hopf3}, in terms of $\Psi$ and $\chi$ reads: 

\begin{eqnarray}
    \begin{bmatrix}
        E_x \\
        E_y
    \end{bmatrix} = 
    \sqrt{\frac{1}{\alpha^2 + \beta^2}} 
    \begin{bmatrix}
        \alpha \\
        \beta
    \end{bmatrix}\\
    \alpha = \cos{\Psi}\cos{\chi} + i\sin{\Psi}\sin{\chi}\\
    \beta = \sin{\Psi}\cos{\chi} - icos{\Psi}\sin{\chi}.
\end{eqnarray}

\bibliographystyle{apsrev}
\bibliography{manuscript.bib} 

\end{document}